%% file: PIMRC_Shah.tex
\documentclass[conference]{IEEEtran}
\usepackage[letterpaper, left=1in, right=1in, bottom=0.9in, top=0.75in]{geometry}
\usepackage{amsmath,amssymb,graphicx,epstopdf,enumitem,pgfplots,cite}
\usepackage{algorithm,algorithmic}
\usepackage[nocomma]{optidef}
\floatname{algorithm}{\small Algorithm}
\usepackage[caption=false,font=footnotesize]{subfig}
\usepackage{balance}
\usetikzlibrary{shapes,arrows,positioning,patterns}

\input{./macros.tex}

\newcommand{\algorithmicbreak}{\textbf{break}}
\newcommand{\BREAK}{\STATE \algorithmicbreak}

\makeatletter
\def\footnoterule{\relax%
  \kern-5pt
  \hbox to \columnwidth{\vrule width 0.5\columnwidth height 0.4pt\hfill}
  \kern4.6pt}
\makeatother

\newcommand\blfootnote[1]{%
  \begingroup
  \renewcommand\thefootnote{}\footnote{#1}%
  \addtocounter{footnote}{-1}%
  \endgroup
}

\newenvironment{varalgorithm}[1]
  {\algorithm[tb!]\renewcommand{\thealgorithm}{#1}}
  {\endalgorithm}

\newlength\figureheight
\newlength\figurewidth

\newcommand\notype[1]{\unskip}

\begin{document}

\title{Association of Networked Flying Platforms with Small Cells for Network Centric\\5G+ C-RAN}

\author{\IEEEauthorblockN{Syed A. W. Shah\IEEEauthorrefmark{1},
Tamer Khattab\IEEEauthorrefmark{1},
Muhammad Zeeshan Shakir\IEEEauthorrefmark{2},
Mazen O. Hasna\IEEEauthorrefmark{1}}
\IEEEauthorblockA{\IEEEauthorrefmark{1}Department of Electrical Engineering, Qatar University, Doha, Qatar\\ Emails: \{syed.shah, tkhattab, hasna\}@qu.edu.qa}
\IEEEauthorblockA{\IEEEauthorrefmark{2}School of Engineering and Computing, University of the West of Scotland, Paisley, Scotland, UK\\
Email: muhammad.shakir@uws.ac.uk}}

\maketitle

\begin{abstract}
5G+ systems expect enhancement in data rate and coverage area under limited power constraint. Such requirements can be fulfilled by the densification of small cells (SCs). However, a major challenge is the management of fronthaul links connected with an ultra dense network of SCs. A cost effective and scalable idea of using network flying platforms (NFPs) is employed here, where the NFPs are used as fronthaul hubs that connect the SCs to the core network. The association problem of NFPs and SCs is formulated considering a number of practical constraints such as backhaul data rate limit, maximum supported links and bandwidth by NFPs and quality of service requirement of the system. The network centric case of the system is considered that aims to maximize the number of associated SCs without any biasing, i.e., no preference for high priority SCs. Then, two new efficient greedy algorithms are designed to solve the presented association problem. Numerical results show a favorable performance of our proposed methods in comparison to exhaustive search.
\end{abstract}

\begin{IEEEkeywords}
drones, small cells, UAVs, network flying platforms, 5G, backhaul/fronthaul network, C-RAN
\end{IEEEkeywords}

\section{Introduction}\label{sec:Intro}
\blfootnote{This publication was made possible by the sponsorship agreement in support of research and collaboration by Ooredoo, Doha, Qatar. The statements made herein are solely the responsibility of the authors.}
Technological advancement and societal development compel the limits of a wireless communication system. Rapid increase in mobile devices, advancement in video services and latest applications such as virtual and augmented reality require enhancement in wireless data rate along with limited power and a widespread coverage. To satisfy these demands, a paradigm shift is needed in the wireless system. Thus, fifth generation and beyond (5G+) systems are supposed to include architectural changes with the use of latest technologies.

For 5G+ systems, an ultra dense network of small cells (SCs) along with milimitere-Wave (mmWave) and free space optics (FSO) is capable of providing hundreds of megahertz of bandwidth under usual power constraints along with widespread coverage \cite{5G_Andrew}. However, such networks impose backhaul traffic limitations as it is difficult to manage a highly dense network of links \cite{robson2012backhaul}. To overcome this limitation, a centralized random access network (C-RAN) can be utilized, where a fronthaul link shares the traffic of the backhaul link \cite{peng2015CRAN}.

An intelligent management is necessary for the fronthaul links due to the ultra dense network of SCs. A wired fiber network is not the best choice due to its high capital expenditure (CAPEX) \cite{CAPEX_opt}. Wireless mmWave/FSO links due to their short range communication require hub points to carry the fronthaul traffic. In urban areas, ground fronthaul hubs are also not a good choice due to the unavailability of a large number of ground locations and non light of sight (NLoS) link losses. Thus, a scalable idea was presented in \cite{ShakirFSOMAG} to utilize the network flying platforms (NFPs) as aerial hubs. Such NFPs includes drones and unmanned aerial vehicles (UAVs) that can be used to communicate wireless information. These NFPs can provide line of signt (LoS) wireless connection as they hover at an altitude of a few hundred meters up to 20 kms. Such a fronthaul network is suitable for many other scenarios apart from a normal LoS wireless connection such as in a number of critical situations (e.g., earthquakes), crowded events (e.g., FIFA world cup) and remote areas (e.g., for mountain climbers). In this work, such NFPs are used to provide fronthaul connectivity to the SCs and efficient greedy algorithms are designed to solve the association problem of the NFPs and SCs.

\subsection{Related Work}\label{sec:RelWork}
A widely used air to ground (ATG) prorogation model was presented in \cite{ATGmodel} that considers aerial communication between NFPs and ground nodes. The coverage area of a single NFP was analytically derived in \cite{ATG_optDrone1}. The maximum coverage area for the case of two NFPs was computed in \cite{TwoDrones} by solving an optimization problem considering their distance and height as an optimization parameters. The association and placement problems of NFPs were targeted in a few articles \cite{IremOneDrone, ElhamBackhaul, ElhamMultiPSO}. The placement of a single NFP used as a base station (BS) was studied in \cite{IremOneDrone} for various urban environments considering only signal to noise ratio (SNR) constraint. Whereas, \cite{ElhamBackhaul} solves the same problem considering backhaul data rate and NFP bandwidth limits. Both \cite{IremOneDrone} and \cite{ElhamBackhaul} solve the placement problem using an exhaustive search. \cite{ElhamMultiPSO} solved the association of multiple NFP-BSs with users using particle swarm optimization (PSO) algorithm considering only signal to interference and noise ratio (SINR). All of the above mentioned works use NFPs as BSs, however, \cite{ShakirFSOMAG} presented the idea of utilizing NFPs as hub points for the SCs but have not designed/solved the related association and placement problem. Same idea was used in \cite{shahNFP} that formulated and solved the association problem of NFP-hubs and SCs using a simple greedy algorithm.
\begin{figure}[tb!]\centering
    \includegraphics[width=8cm]{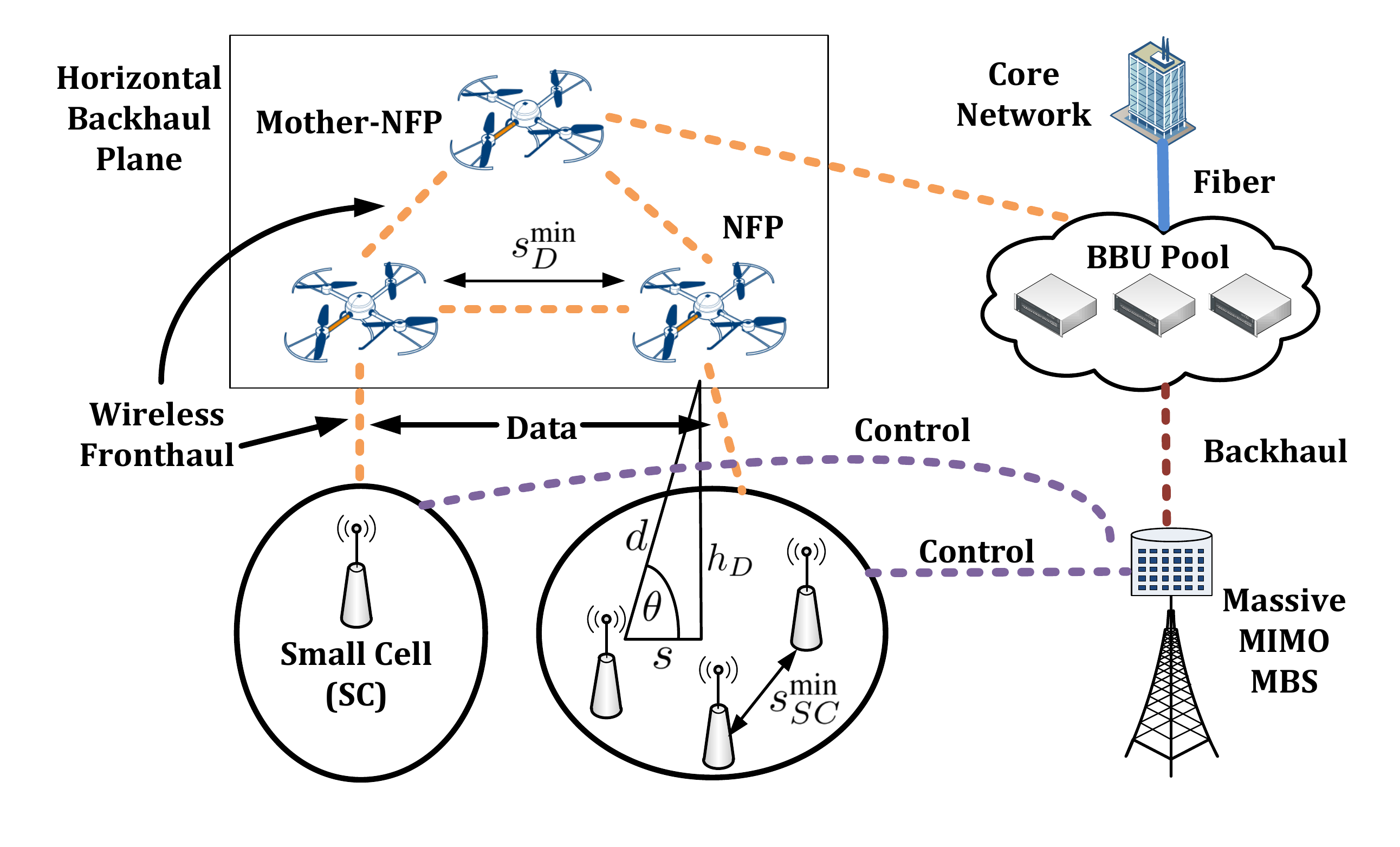}
	\caption{Graphical illustration of NFPs and SCs in a 5G+ C-RAN.}
	\label{fig:SysMod}
\end{figure}

\subsection{Our Contributions}\label{sec:OurCont}
This work considers the association problem of NFPs and SCs for the network centric case i.e., from network point of view where the network's objective is to serve maximum number of SCs without any biasing. A related work was presented for the user centric case in \cite{shahNFP}, where there was a biasing due to sum data rate i.e., the SCs demanding a high data rate were given priority to get associated first in order to increase the total sum data rate. Similar to the work in \cite{shahNFP}, same practical constraints such as maximum backhaul data rate, maximum number of transceivers that NFP can carry to maintain links with SCs, maximum bandwidth limit of NFP and interference between neighbouring SCs are considered that have not been used before in any other work. Further, a practical stochastic geometry approach is used for the distribution of SCs and NFPs by maintaining a distance between their neighbours. Two simple but efficient greedy algorithms are presented for two scenarios including i) when SCs have limited processing power then a centralized algorithm is used named as Centralized Maximal Cells Algorithm (CMCA) and ii) when SCs have enough processing power, then a faster distributed algorithm can be used that named as Distributed Maximal Cells Algorithm (DMCA). A performance comparison via numerical results shows a similar association and more than 80\% faster run time speed of the proposed algorithms in terms of exhaustive branch and bound (B\&B) method.

The remainder of the article is organized as follows. Section \ref{sec:SysMod} consists of the system model and the association problem of NFPs and SCs for the network centric case. Section \ref{sec:OptAlgo} presents the proposed algorithms to solve the association problem. Numerical results and related discussions are presented in Section \ref{sec:SimRes} and finally Section \ref{sec:Conc} concludes the paper.

\section{System Model and Problem Formulation}\label{sec:SysMod}
Consider a C-RAN architecture of 5G+ system as shown in Fig. \ref{fig:SysMod} that consists of an ultra dense network of SCs, a macrocell base station (MBS) and a core network. SCs consists of a remote radio head (RRH) and may or may not have a processing unit depending on the architecture \cite{peng2015CRAN}. Mainly, the processing of control and data information is handled by a centralized base band unit (BBU) pool consisting of a number of BBUs equal to the number of cell sites. The control information of users flows through the backhaul network including connections between SCs and core network via MBS and BBU pool. The MBS consists of a massive multiple-input multiple-output (MIMO) antenna that connects with SCs and BBU pool using mmWave links, then this BBU pool process the information and transfers it to the core network through a fiber optic network.

Instead of a usual wired fronthaul network, here NFPs are used along with mmWave/FSO to route the data traffic of users between SCs and the core network via BBU pool. The NFPs have a two level hierarchy, where a number of NFPs are spread over a horizontal plane of height $h_D$ from ground level. They are connected to each other through FSO links and with SCs through mmWave links, where the FSO link losses are not considered in this study. The NFPs act as hub points for carrying traffic between SCs and core network but for brevity they are referred as NFPs instead of NFP-hubs. This layer of NFPs is also connected via FSO link with mother-NFP at a height greater than $h_D$. These NFPs are only allowed to share control signals with each other, where each NFP should directly route the data traffic to the mother-NFP. Such control signals includes the SINR of SCs and NFPs links, bandwidth and data rate requirements of the SCs. In this work, only those SCs are considered that are active during time interval $\begin{bmatrix} 0 & T \end{bmatrix}$ and it is assumed that the system does not change during this time duration. Before formulating the problem, first the air-to-ground (ATG) path loss model is discussed below.

\subsection{Air-to-Ground Path Loss Model}\label{sec:ATG_Model}
A widely used ATG path loss model presented in \cite{ATGmodel} and \cite{ATG_optDrone1} is adopted here. This model is based on two propagation groups namely: i) LoS receivers and ii) NLoS receivers. The probability of LoS $P(\text{LoS})$ depends on the considered environment (such as rural, urban, or others) and the orientation of NFPs and ground SCs and it was formulated in \cite{ATGmodel} and \cite{ATG_optDrone1} as
\begin{equation}\label{Prob_LoS}
    P(\text{LoS}) = \frac{1}{1 + \alpha \exp \left\{ -\beta \left( \frac{180}{\pi} \theta - \alpha \right) \right\}}
\end{equation}
where $\alpha$ and $\beta$ are constant values that depend on the specific environment. Here, $\theta = \arctan \left( \frac{h_D}{s} \right)$ represents the elevation angle from the ground SC to the NFP, where $s = \sqrt{ \left( x - x_D \right)^2 + \left( y - y_D \right)^2 }$ denotes the horizontal distance between them. The positions of SCs and NFPs in a cartesian coordinate system with respect to the origin are denoted by $\left(x,y\right)$ and $\left(x_D,y_D,h_D\right)$, respectively. This model ignores the random variations in the channel and presents the average path loss as
\begin{equation}\label{PathLoss}
  \begin{aligned}
    \text{PL}(dB) = &10 \log \left( \frac{4 \pi f_c d}{c} \right)^\gamma + P(\text{LoS}) \eta_{\text{LoS}}\\
                    &+ P(\text{NLoS}) \eta_{\text{NLoS}}
  \end{aligned}
\end{equation}
where the first term represents free space path loss (FSPL) that depends on carrier frequency $f_c$, speed of light $c$, PL exponent $\gamma$ and the distance $d=\sqrt{h_D^2+s^2}$ between NFP and SC. Variables $\eta_{\text{LoS}}$ and $\eta_{\text{NLoS}}$ represent additional losses for LoS and NLoS links, respectively and $P(\text{NLoS}) = 1 - P(\text{LoS})$. All parameters in \eqref{PathLoss} depend on the environment.

\subsection{Problem Formulation}\label{sec:Prob_Form}
Consider a transmission of user data between $N_{SC}$ small cells and the core network via fronthaul link consisting of $N_D$ NFPs except the mother-NFP. Considering a stochastic geometry approach, both SCs and NFPs are distributed randomly in a square region of area $A$ using \emph{Matern} type-I hard-core process \cite{matern2013} with an average density of $\lambda$ per $\text{m}^2$ having a minimum separation of $s_{SC}^{\text{min}}$ and $s_D^{\text{min}}$ with their neighbors, respectively. This provides a random distribution points of SCs and NFPs denoted as $\left(x_i,\,y_i\right)$ and $(x_{D_j}, y_{D_j}, h_{D_j})$, respectively, where $i \in \left\{ 1, \ldots, N_{SC}  \right\}$ and $j \in \left\{ 1, \ldots, N_D  \right\}$.

The communication between the core network and SCs is limited by a number of factors that also affects the association of NFPs and SCs. Thus, depending on these factors, out of all the available pairs of NFPs and SCs only few can be associated. First of all, the following discussion presents the limiting factors of the system and then formulate the association problem of NFPs and SCs based on those factors.

The backhaul link running from the core network to the mother-NFP via BBU pool limits the maximum allowed data rate of the network, that is referred here as backhaul data rate $R$. This means that the sum of the data rate from all the NFP and SC pairs cannot exceed the backhaul data rate $R$. Let us denote the requested data rate of SC $i$ associated with NFP $j$ by $r_{ij}$, then this constraint can be written as \eqref{cons1} where $A_{ij}$ is an entry of ($N_{SC} \times N_D$) association matrix $\Am$ that shows the association of SCs and NFPs as
\begin{equation}\label{Assoc_Mat}
      A_{ij} =
        \left\{
          \begin{array}{ll}
            1, & \hbox{if SC $i$ is connected with NFP $j$,} \\
            0, & \hbox{otherwise.}
          \end{array}
        \right.
\end{equation}

The next limit is posed by the frontlink FSO link in the next hop, i.e., from mother-NFP to each NFP. Depending upon the quality of FSO link, each NFP $j$ is allocated a maximum bandwidth $B_j$ that can be distributed among associated SCs. This limits the sum of requested bandwidth of all SCs associated with NFP $j$ and it can be mathematically represented as \eqref{cons2}. The allocated bandwidth $b_{ij} = \frac{r_{ij}}{\eta_{ij}}$ of SC $i$ and NFP $j$ pair depends on $r_{ij}$ and spectral efficiency $\eta_{ij} =$ $\log_2 \left( 1+\text{SINR}_{ij} \right)$, where SINR can be expressed as
\begin{equation}\label{eq:SINR}
      \text{SINR}_{ik} = \frac{P_{r_{ik}}}{\sum_{j=1,j \neq k}^{N_D} P_{r_{ij}} + \sigma}
\end{equation}
Here, $P_{r_{ij}}$ represents the received power from NFP $j$ to the SC $i$ and $\sigma$ represents the noise floor of the link.

In the next hop, i.e., from NFP $j$ to SC $i$, the mmWave fronthaul link should satisfy a quality of service (QoS) requirement. Every NFP can serve SCs placed inside a specific area computed using \eqref{PathLoss} for fixed positions of NFPs, SCs and a maximum path loss \cite{ATG_optDrone1, TwoDrones}. This maximum path loss is related to minimum SINR that is required to serve a SC via mmWave link. Thus, each NFP SC pair link should satisfy a minimum SINR QoS requirement which can be written as \eqref{cons3}.

Considering all the above mentioned constraints of the 5G+ C-RAN network, for fixed positions of NFPs and SCs, we search for the best possible association between them. The objective of the network centric case is to serve the maximum number of SCs. Such a problem can be formulated as
\begin{subequations}
\label{eq:Opt_Prob}
\begin{alignat}{3}
    \max_{\{A_{ij}\}} \quad \sum_{i=1}^{N_{SC}} &\sum_{j=1}^{N_D} \cdot A_{ij} \label{Obj_Fun}\\
    \intertext{subject to}
    \sum_{i=1}^{N_{SC}} \sum_{j=1}^{N_D} r_{ij} \cdot A_{ij} \; &\leq \; R \label{cons1}\\
    \sum_{i=1}^{N_{SC}} b_{ij} \cdot A_{ij} \; &\leq \; B_j, && \quad \forall j \label{cons2}\\
    \text{SINR}_{ij} \cdot A_{ij} \; &\geq \; \text{SINR}_{\text{min}}, && \quad \forall i,j \label{cons3}\\
    \sum_{i=1}^{N_{SC}} A_{ij} \; &\leq \; N_{l_j}, && \quad \forall j \label{cons4}\\
    \sum_{j=1}^{N_D} A_{ij} \; &\leq \; 1, && \quad \forall i \label{cons5}
\end{alignat}
\end{subequations}
Constraint \eqref{cons4} shows that NFP $j$ can maintain a maximum of $N_{l_j}$ links with SCs as per the number of transceivers. Further, each SC can be associated to a maximum of one NFP that is included in constraint \eqref{cons5}.

\section{Optimization Algorithms}\label{sec:OptAlgo}
The optimization problem in \eqref{eq:Opt_Prob} is a binary linear integer program (BILP) and it can be easily noticed that this is an NP-hard problem \cite{NPhard2}. It is well known that there exits no standard method to solve such an optimization problem. Therefore, we present here two simple but efficient greedy solutions that are designed to solve the optimization problem \eqref{eq:Opt_Prob}. One of the algorithm is applicable for the architecture where SCs lack the processing power, thus, algorithm works in a centralized manner and named as Centralized Maximal Cells Algorithm (CMCA). The other presented method works in a distributed fashion for the case where processing power of SCs and NFPs is utilized too and is named as Distributed Maximal Cells Algorithm (DMCA). For the performance comparison an exhaustive search method known as branch and bound (B\&B) method \cite{BandB_Algo} is utilized as an optimal benchmark solution.

To setup the system model, initialization algorithm presented in \cite{shahNFP} is utilized. It is assumed that all the NFPs are symmetric, i.e., $h_{D_j}=h_D$, $B_j=B$ and $N_{l_j}=N_l$, however, the presented algorithms are applicable for the general case of optimization problem \eqref{eq:Opt_Prob} with necessary modifications. Using initialization algorithm, the NFPs and SCs are distributed randomly in a square region of area $A$. At this point, a snapshot of NFPs and SCs is obtained which provides their respective numbers $N_D$ and $N_{SC}$ and positions which combined with data rate requirements of SCs are used to compute the bandwidth and SINR parameters. Then, this information is passed to the proposed algorithms to find the optimum association between NFPs and SCs.

\subsection{Proposed Distributed Greedy Algorithm}\label{sec:DistAlgo}
This algorithm is divided into four steps that are distributed among SCs (first step), NFPs (second step) and mother-NFP (last two steps) to utilize their processing power and to speed up the optimization process. At every step, the algorithm takes care of either one or two constraints of the optimization problem \eqref{eq:Opt_Prob}.

\subsubsection{\textbf{Step 1}}\label{sec:Step1Sol}
Using the $\text{SINR}_{ij}$ from \eqref{eq:SINR}, every SC $i$ compares its SINR with the minimum SINR requirement for all the available NFPs satisfying constraint \eqref{cons3}. SC $i$ then selects only one NFP having minimum sum bandwidth and data rate, i.e., $\min (b_{ij} + r_{ij})$ out of the other NFPs that satisfy constraint \eqref{cons3}. Thus, at this step algorithm takes care of the constraints \eqref{cons3} and \eqref{cons5}.

\subsubsection{\textbf{Step 2}}\label{sec:Step2Sol}
Each NFP $j$ receive a number of association requests from SCs. Out of these requests, every NFP $j$ performs action on its own received list and selects a maximum of $N_l$ requests satisfying the constraint \eqref{cons4}. The selection of SC is based on the minimum sum of bandwidth $b_{ij}$ and data rate $r_{ij}$, thus trying to maximize the number of SCs as per the objective criterion \eqref{Obj_Fun} considering constraints \eqref{cons1} and \eqref{cons2}. Thus, NFP $j$ first selects the SC having minimum sum bandwidth and data rate, i.e., $\min (b_{ij} + r_{ij})$ out of its list and then selects the next SC with a higher value. NFP $j$ keeps track of the number of links and bandwidths of the associated SCs using counters $C_{N_l}^j$ and $C_b^j$, respectively. Furthermore, before associating the SC $i$ with NFP $j$, NFP validates the constraint \eqref{cons2} to ensure that each NFP $j$ does not exceed its maximum bandwidth limit $B$, i.e., $C_b^j + b_{ij} \leq B$ and then after the association updates the respective counters. The process completes if NFP $j$ reaches either of the maximum limits including the number of links $N_l$ and bandwidth $B$ or if the number of requests of SCs for NFP $j$ ends.

This step is designed in a way so that it can be performed in parallel at every NFP to speed up the association process and to distribute the processing load among all the NFPs. It is to be noted that this step uses information of constraint \eqref{cons1}, however, as this step is being performed in parallel thus it does not keep track of this constraint. Thus, at the end the association matrix $\Am$ contains a maximum of $N_l$ ones at every column $j$ and constraints \eqref{cons2} and \eqref{cons4} are satisfied.

All the NFPs pass their information to the mother-NFP; this information includes their association decisions and the respective bandwidth, data rate and SINR. A counter $C_r$ having information of the total sum data rate of the associated SCs is initialized.
\subsubsection{\textbf{Step 3}}\label{sec:Step3Sol}
This step is performed if the backhaul data rate limit is not reached, i.e., $C_r < R$. Mother-NFP goes through the association matrix $\Am$ to find the unassociated SCs and also computes the remaining available resources, i.e., number of links and bandwidth for each NFP using counter $C_{N_l}^j$ and $C_b^j$, respectively. Mother-NFP requests the bandwidth $b_{ij}$ and data rate $r_{ij}$ parameters of the links between unassociated SCs and NFPs that satisfy the SINR constraint \eqref{cons3} at step 1 of this algorithm. Then, mother-NFP selects the NFP to SC link that provides the minimum sum bandwidth and data rate, i.e., $\min (b_{ij} + r_{ij})$ out of all the possible unassociated SCs and NFPs pairs. It associates the selected pair after verifying the constraints of $N_l$ \eqref{cons4} and $B$ \eqref{cons2} for the selected NFP $j$ and $R$ \eqref{cons1} for all the NFPs. In this way, it keeps associating the remaining SCs until the resources are available or all the SCs get associated.

\subsubsection{\textbf{Step 4}}\label{sec:Step4Sol}
This step is performed only if the backhaul data rate limit is exceeded, i.e., $C_r > R$, otherwise the algorithm completes. So, at this step, mother-NFP takes care of the data rate constraint \eqref{cons1} using counter $C_r$. Mother-NFP dissociates some of the SCs in the following manner. It searches for the SC having maximum data rate $r_{ij}$ and dissociates it if $C_r - r_{ij} \geq R$, otherwise, selects the SC with next lower data rate. After every disassociation, mother-NFP verifies the data rate constraint \eqref{cons1} such that if $C_r \leq R$ then the algorithms completes, otherwise repeats the same procedure.

This algorithm provides an efficient solution of the optimization problem \eqref{eq:Opt_Prob} in four simple steps and is summarized in Algorithm \ref{Dist_Algo}.

\begin{varalgorithm}{DMCA}
\small
\caption{\small{Distributed Maximal Cells Algorithm}}
\label{Dist_Algo}
\begin{algorithmic}[1]
    \REQUIRE $N_{SC}, \; N_D, \; N_l, \; \text{SINR}_{\text{min}}, \; B, \; R, \; \text{SINR}_{ij}, \; r_{ij}, \; b_{ij}$
    \ENSURE $\Am$
    \STATE Initialize: $\Am = \emptyset$
    \STATE \textbf{Step 1:}
    \FOR {$i = 1$ \TO $N_{SC}$}
        \STATE Make a list of NFPs that satify $\text{SINR}_{\text{min}}$ criterion
        \STATE Out of the list send association request to NFP $j$ such that $\min (b_{ij} + r_{ij})$
    \ENDFOR
    \STATE \textbf{Step 2:}
    \STATE Initialize counters: $C_{N_l}^j = 0$, $C_b^j = 0$ $\forall j$
    \FOR {$j = 1$ \TO $N_D$}
        \WHILE {$C_{N_l}^j < N_l$ $\wedge$ $C_b^j < B$}
            \STATE Find SC $i$ with $\min (b_{ij} + r_{ij})$
            \IF {$C_b^j + b_{ij} \leq B$}
                \STATE Update $A_{ij}=1$, $C_{N_l}^j = C_{N_l}^j+1$ and $C_b^j = C_b^j + b_{ij}$
            \ELSE
                \BREAK
            \ENDIF
        \ENDWHILE
    \ENDFOR
    \STATE Initialize: $C_r$ as total data rate of associated SBSs
    \STATE \textbf{Step 3:}
        \WHILE {$C_r < R$}
            \STATE Find unassociated SC by scanning matrix $\Am$
            \STATE Find NFPs with remaining resources using $C_{N_l}^j$ and $C_b^j$
            \STATE Associate NFP SC pair which gives $\min (b_{ij} + r_{ij})$
        \ENDWHILE
    \STATE \textbf{Step 4:}
        \WHILE {$C_r > R$}
            \STATE Select NFP SC pair with $\max (b_{ij} + r_{ij})$
            \STATE De-associate selected NFP to SC pair as $A_{ij} = 0$
            \STATE Update total data rate as $C_r = C_r - r_{ij}$
        \ENDWHILE
\end{algorithmic}
\end{varalgorithm}

\subsection{Proposed Centralized Greedy Algorithm}\label{sec:CentAlgo}
This centralized algorithm having slower run time speed than \ref{Dist_Algo} is suitable for the C-RAN architecture where the BBU pool does all the processing. It can be implemented at either BBU pool or mother-NFP that can obtain the necessary control information from the BBU pool. Similar to Algorithm \ref{Dist_Algo}, this centralized algorithm is designed on the same strategy that is to select the SCs with minimum sum bandwidth and data rate in order to accommodate more and more SCs in the NFP bandwidth $B$ and backhaul data rate $R$ limits.

\begin{varalgorithm}{CMCA}
\small
\caption{\small{Centralized Maximal Cells Algorithm}}
\label{Cent_Algo}
\begin{algorithmic}[1]
    \REQUIRE $N_{SC}, \; N_D, \; N_l, \; \text{SINR}_{\text{min}}, \; B, \; R, \; \text{SINR}_{ij}, \; r_{ij}, \; b_{ij}$
    \ENSURE $\Am$
    \STATE Make a list of NFP to SC links that satify $\text{SINR}_{\text{min}}$ criterion
    \STATE Initialize counters: $C_{N_l}^j = 0$, $C_b^j = 0$ $\forall j$ and $C_r = 0$
    \WHILE {List of NFP to SC links is not empty}
        \STATE Find SC $i$ and NFP $j$ pair with $\min (b_{ij} + r_{ij})$
        \IF {$C_r + r_{ij} \leq R$}
            \IF {$C_b^j + b_{ij} \leq B$  $\wedge$ $C_{N_l}^j < N_l$}
                \STATE Update $A_{ij}=1$, $C_{N_l}^j = C_{N_l}^j+1$, $C_b^j = C_b^j + b_{ij}$ and $C_r = C_r + r_{ij}$
                \STATE Remove other links of selected SC $i$ from the list
            \ELSE
                \STATE Remove all links of NFP $j$ from the list
            \ENDIF
        \ELSE
            \BREAK
        \ENDIF
    \ENDWHILE
\end{algorithmic}
\end{varalgorithm}

The algorithm starts by generating a list of NFP to SC links that satisfy the SINR constraint in \eqref{cons3}. Also, it initializes the three counters, i.e., the number of links $C_{N_l}^j$, bandwidth $C_b^j$ of NFP and sum data rate $C_r$ of all NFPs. Out of the list of NFP to SC links, the link that provides minimum sum bandwidth and data rate, i.e., $\min (b_{ij} + r_{ij})$ is chosen. Now, algorithm verifies the constraints of data rate \eqref{cons1}, bandwidth \eqref{cons2} and number of links \eqref{cons4} such that $C_r + r_{ij} \leq R$, $C_b^j + b_{ij} \leq B$ and $C_{N_l}^j + 1 \leq N_l$, respectively. If the selected NFP SC pair pass the verification stage then they are associated to each other and association matrix $\Am$ is updated along with all three counters. If the data rate constraint is not passed then the process is terminated as it means the association has reached the backhaul data rate limit. In case, either of the bandwidth or the number of links constraints are not passed then all the links of the selected NFP $j$ are removed from the list as it means this NFP $j$ cannot accommodate any more SCs. Further, during association of SC $i$, all the other possible links of the selected SC are removed from the list of NFP to SC links in order to satisfy the constraint \eqref{cons5} that restricts a SC to be linked with a maximum of one NFP. The process is repeated until the list ends or the resources ends that can be tracked using the three counters. The steps are summarized in Algorithm \ref{Cent_Algo}.

\section{Numerical Results}\label{sec:SimRes}
Consider a C-RAN 5G+ system as shown in Fig. \ref{fig:SysMod}, where the SCs and NFPs are distributed randomly in a square region of area $A = 16$ km$^2$. Both the SCs and NFPs are distributed using \emph{Matern} type 1 process with same density $\lambda$. A minimum separation of $s_{SC}^{\text{min}} = 300$ meters is maintained between neighbouring SCs. However, among neighbouring NFPs the minimum distance $s_D^{\text{min}}$ is computed using \eqref{PathLoss} considering a maximum path loss $\text{PL}_{\text{max}}$. Then, the data rate is randomly assigned to the SCs from a pre-defined vector $\rv_{\text{SBS}}$. Note that, here it is assumed that a SC $i$ will demand the same data rate from every NFP i.e., $r_{ij} = r_i, \forall j$. Considering this and the other parameters defined in Table \ref{tab:SimPar}, the parameters for every NFP to SC pair links are calculated such as $b_{ij}$ and $\text{SINR}_{ij}$. Finally, the necessary parameters are passed to the algorithms to find the best possible association between NFPs and SCs by minimizing optimization problem \eqref{eq:Opt_Prob}. In the following, the parameters remain the same unless otherwise stated, however a number of scenarios are considered i.e., different random distributions of NFPs and SCs in the same square region.

\begin{table}[tb!]
\renewcommand{\arraystretch}{1.3}
\centering
\caption{Simulation Parameters}
\label{tab:SimPar}
\begin{tabular}{|c|c|c|c|}
\hline
\textbf{Parameter}          & \textbf{Value}        & \textbf{Parameter}       & \textbf{Value}   \\ \hline
$\alpha$                    & 9.61                  & $\beta$                  & 0.16             \\ \hline
$\eta_{\text{LoS}}$         & 1 dB                  & $\eta_{\text{NLoS}}$     & 20 dB            \\ \hline
$f_c$                       & 2 GHz                 & $P_t$                    & 5 Watts          \\ \hline
$\text{SINR}_{\text{min}}$  & -5 dB                 & $\text{PL}_{\text{max}}$ & 115 dB           \\ \hline
$\lambda$                   & 5 $\times 10^{-6}$    & $h_{D_{\text{max}}}$     & 300 meters       \\ \hline
$N_{SC}$                    & 30                    & $N_D$                    & 3                \\ \hline
$\rv_{\text{SBS}}$            & \multicolumn{3}{c|}{ \{ 30, 60, 90, 120, 150 \} Mbps }       \\ \hline
\end{tabular}
\end{table}

Fig. \ref{fig:fig1} shows one of the considered scenarios for the distribution of NFPs and SCs, where only 2D aerial view is shown as the height of NFPs is same. Also, in this case the association is same for both Algorithms \ref{Dist_Algo} and \ref{Cent_Algo}, therefore, their results are shown jointly in Fig. \ref{fig:fig1b}. It can be noticed by comparing Fig. \ref{fig:fig1a} and \ref{fig:fig1b} that B\&B and our proposed algorithms associates 28 and 27 SCs, respectively. The performance is close but the difference is mainly because of the backhaul data rate constraint. For this scenario, as per the rate distribution the sum data rate of total 30 SCs is 3.18 Gbps which exceeds the considered data rate constraint of $R = 2.9$ Gbps. Therefore, B\&B method only associates 28 SCs as it drops 2 SCs requesting a data rate of 0.15 Gbps each. However, our proposed algorithms are based on combined minimum sum bandwidth and data rate instead of exhaustive search of B\&B method, so they drop 3 SCs with data rate of 0.15 Gbps and 2 with 0.12 Gbps. It happens because the second SC dropped by B\&B method requires a bandwidth such that it results in a sum bandwidth and data rate lower than the ones with 0.12 Gbps data rate. If our algorithms have a different weighting criterion between bandwidth and data rate, they would have dropped 2 SCs giving the same results as of B\&B method. However, this weighting criterion suits this scenario only and it changes for every other scenario, that is why in the proposed algorithms, we selected the simplest weighting criterion for minimum sum bandwidth and data rate.

\begin{figure}[tb!]\centering
	\setlength\figureheight{3.7cm}
	\setlength\figurewidth{3.7cm}
    \subfloat[B\&B association.]{\label{fig:fig1a}
	\input{./Fig1a.tikz}} \vfill \medskip
    \subfloat[Algorithm \ref{Dist_Algo} and \ref{Cent_Algo} association.]{\label{fig:fig1b}
	\input{./Fig1b.tikz}}
	\caption{2D view of a random distribution and association of NFPs and SCs with constraints $N_l = 16$, $B = 1$ GHz, $R = 2.9$ Gbps.}
	\label{fig:fig1}
\end{figure}
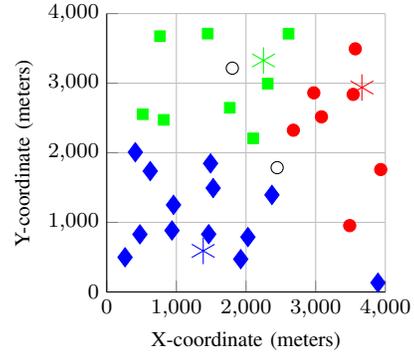

\begin{figure}[tb!]\centering
	\setlength\figureheight{3.5cm}
	\setlength\figurewidth{6.5cm}
	\input{./Fig2.tikz}
	\caption{\% of unassociated SCs vs $R_r$ for 3 algorithms with restrictions $B = 2$ Gbps, $N_l = 30$ averaged over 1000 different scenarios.}
	\label{fig:fig2}
\end{figure}
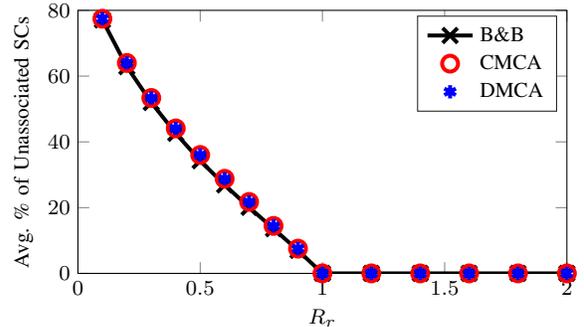

Fig. \ref{fig:fig2} plots the average percentage of unassociated SCs vs. $R_r$ which is the ratio of the backhaul data rate limit $R$ to the sum data rate of the $N_{SC}$ SCs. 1000 different scenarios are considered and for each scenario the number of unassociated SCs are used to compute the percentage that is then averaged over 1000 scenarios. For all the scenarios, the same ratio $R_r$ is maintained by varying the backhaul data rate limit $R$ and total sum data rate for each scenario. This figure highlights the effect of the backhaul data rate constraint \eqref{cons1} on the association of SCs. The percentage of unassociated SCs decreases with the increase in the ratio $R_r$ until it reaches 1 and remains zero for $R_r \geq 1$, i.e., all SCs gets associated if resources are equal or greater than the requirements. Further, it can be seen that the performance of our proposed algorithms is same as of the optimal solution by B\&B method.

\begin{figure}[tb!]\centering
	\setlength\figureheight{3.5cm}
	\setlength\figurewidth{6.5cm}
	\input{./Fig3.tikz}
	\caption{\% of unassociated SCs vs $R_b$ for 3 algorithms with constraints $R = 5$ Gbps, $N_l = 30$ averaged over 1000 different scenarios.}
	\label{fig:fig3}
\end{figure}
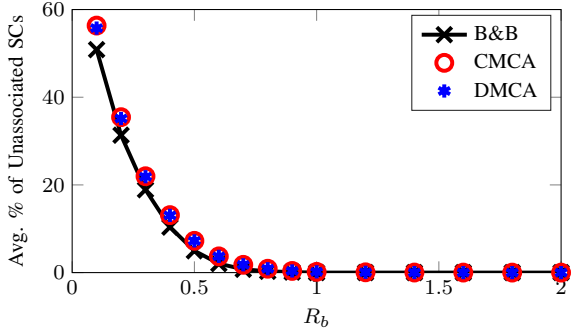

Fig. \ref{fig:fig3} depicts the average percentage of unassociated SCs vs. $R_b$ which is the ratio of the bandwidth limit of NFPs $B$ to the sum bandwidth of the associated SCs with one NFP. As symmetric NFPs are considered in this work so the ratio $R_b$ is computed using the NFP where associated SCs demand maximum sum bandwidth among other NFPs and the $B$ is varied also to maintain the same $R_b$ for all scenarios. For more tight restriction of $B$, i.e., constraint \eqref{cons2}, the proposed algorithms deviates more from the optimal result. The effect of bandwidth limit of NFPs $B$ on the association problem is more non-linear than the backhaul data rate limit $R$. This shows that in the strategy of the proposed algorithms, i.e., $\min (b_{ij} + r_{ij})$, bandwidth should be weighted more than the data rate.

\begin{figure}[tb!]\centering
	\setlength\figureheight{3cm}
	\setlength\figurewidth{6.5cm}
	\input{./Fig4_new.tikz}
	\caption{Run time for all three algorithms vs. 30 different scenarios for constraints $R = 2.9$ Gbps, $B = 1$ GHz and $N_l = 16$.}
	\label{fig:fig4}
\end{figure}
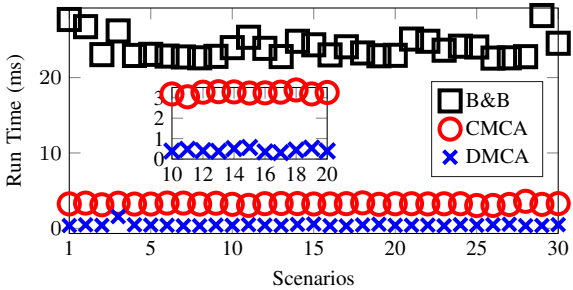
\balance
Fig. \ref{fig:fig4} compares the run time speed of the proposed algorithms with the B\&B method for 30 different scenarios. In all of the scenarios, the constraints \eqref{cons1} to \eqref{cons5} are considered, so resources are limited. If the results are averaged over 1000 scenarios then the average run time speed of B\&B, \ref{Cent_Algo} and \ref{Dist_Algo} results in 27.2561 ms, 3.4805 ms and 0.4919 ms. This means that algorithms \ref{Cent_Algo} and \ref{Dist_Algo} results in 87.49\% and 98.2\% decrease in run time speed as compared to the B\&B method. This shows that the proposed methods are practically applicable due to their lower complexity, fast speed and same average performance.

\section{Conclusions}\label{sec:Conc}
The idea of employing NFPs to provide fronthaul connectivity to SCs is used in this work. The association problem of NFPs and SCs is formulated considering a number of practical constraints including backhaul data rate limit, maximum number of links and bandwidth limit of NFPs, interference between SC and NFPs using SINR criterion. Two greedy algorithms are presented for different types of C-RAN architecture that tries to maximize the number of SCs. One of the algorithm is presented for the case where SCs have the processing capability, so it utilizes the processing power of SCs, NFPs and mother-NFP and thus named as Distributed Maximal Cells Algorithm (DMCA). Whereas, the other one is designed for the architecture where SCs lacks the processing power thus it works at BBU pool or mother-NFP and named as Centralized Maximal Cells Algorithm (CMCA). Both algorithms can be practically implemented as they have lower complexity as compared to B\&B exhaustive method. Numerical results show that both algorithms have nearly the same performance as of B\&B method.


\bibliographystyle{IEEEtran}
\bibliography{PIMRC_Shah}

\end{document}

%% file: macros.tex
\long\def\comment#1{}

\newfont{\bbb}{msbm10 scaled 700}

\newfont{\bb}{msbm10 scaled 1100}

\newcommand{\rv}{{\bf r}}

\newcommand{\Am}{{\bf A}}



%% file: Fig1a.tikz
\footnotesize
\begin{tikzpicture}
\begin{axis}[
width=\figurewidth,
height=\figureheight,
scale only axis,
xmin=0, xmax=4000,
xlabel near ticks,
xlabel={X-coordinate (meters)},
xmajorgrids,
ymin=0, ymax=4000,
ylabel near ticks,
ylabel={Y-coordinate (meters)},
ymajorgrids,
zmin=0, zmax=400,
ztick={\empty},
zmajorgrids,
view={0}{90},
axis x line*=bottom,
axis y line*=left,
axis z line*=left,
]
\addplot[only marks,mark=o,mark options={},mark size=2.3pt,color=black] plot table[row sep=crcr,]{%
2447.08677348132	1786.3174722349\\
1803.31432580452	3215.65349486806\\
};

\addplot[only marks,mark=*,mark options={},mark size=2.3pt,color=red] plot table[row sep=crcr,]{%
2679.77717419022	2324.0694180278\\
3083.34345576156	2516.60170846886\\
3538.44371941934	2838.83053919693\\
2973.31351951974	2861.08756541379\\
3934.29843689313	1758.93243843101\\
3569.12824938997	3493.56898413911\\
3488.77484308511	953.401066684554\\
};

\addplot3[only marks,mark=asterisk,mark options={},mark size=5pt,color=red] plot table[row sep=crcr,]{%
3663.95781048284	2937.49259770534	300\\
};

\addplot[only marks,mark=square*,mark options={},mark size=2pt,color=green] plot table[row sep=crcr,]{%
517.088168808934	2554.50291903537\\
764.860820586706	3675.18698663574\\
1450.32784687507	3712.02901900719\\
819.31547602297	2472.92441543969\\
2311.86209774807	2992.84465164545\\
2610.98354317796	3711.01622638315\\
2102.95467293641	2206.74691110181\\
1767.42925685483	2646.78000160129\\
};
\addplot3[only marks,mark=asterisk,mark options={},mark size=5pt,color=green] plot table[row sep=crcr,]{%
2251.14726144461	3327.02225494686	300\\
};
\addplot[only marks,mark=diamond*,mark options={},mark size=3.6092pt,color=blue] plot table[row sep=crcr,]{%
959.194207575183	1251.171990357\\
2372.47106526323	1394.30747890552\\
263.011068279709	498.337998101655\\
624.861786845241	1736.9682990966\\
476.470571076389	827.845867787273\\
1924.37118987097	471.623403817135\\
2027.52366578952	789.177430179262\\
1528.73536653258	1492.82624142829\\
1464.32251151287	830.417524337401\\
3894.35169051182	132.948617957397\\
1490.83469637876	1847.25739444395\\
937.079410959819	883.285140724237\\
410.617360335338	2008.13805127995\\
};
\addplot3[only marks,mark=asterisk,mark options={},mark size=5pt,color=blue] plot table[row sep=crcr,]{%
1385.26798535204	590.221976969278	300\\
};
\end{axis}
\end{tikzpicture}%

%% file: Fig1b.tikz
\footnotesize
\begin{tikzpicture}
\begin{axis}[%
width=\figurewidth,
height=\figureheight,
scale only axis,
xmin=0, xmax=4000,
xlabel near ticks,
xlabel={X-coordinate (meters)},
xmajorgrids,
ymin=0, ymax=4000,
ylabel near ticks,
ylabel={Y-coordinate (meters)},
ymajorgrids,
zmin=0, zmax=400,
ztick={\empty},
zmajorgrids,
view={0}{90},
axis x line*=bottom,
axis y line*=left,
axis z line*=left,
legend style={at={(-0.4,-0.4)},legend columns=-1,anchor=south west,legend cell align=left,align=left,draw=white!15!black}
]
\addplot[only marks,mark=o,mark options={},mark size=2.3pt,color=black] plot table[row sep=crcr,]{%
2447.08677348132	1786.3174722349\\
3488.77484308511	953.401066684554\\
3894.35169051182	132.948617957397\\
};
\addlegendentry{Unassociated SCs};

\addplot[only marks,mark=*,mark options={},mark size=2.3pt,color=red] plot table[row sep=crcr,]{%
3083.34345576156	2516.60170846886\\
3538.44371941934	2838.83053919693\\
2973.31351951974	2861.08756541379\\
3934.29843689313	1758.93243843101\\
3569.12824938997	3493.56898413911\\
};
\addlegendentry{Associated SCs};

\addplot3[only marks,mark=asterisk,mark options={},mark size=5pt,color=red] plot table[row sep=crcr,]{%
3663.95781048284	2937.49259770534	300\\
};
\addlegendentry{NFPs};

\addplot[only marks,mark=square*,mark options={},mark size=2pt,color=green] plot table[row sep=crcr,]{%
517.088168808934	2554.50291903537\\
764.860820586706	3675.18698663574\\
1450.32784687507	3712.02901900719\\
2679.77717419022	2324.0694180278\\
819.31547602297	2472.92441543969\\
2311.86209774807	2992.84465164545\\
2610.98354317796	3711.01622638315\\
2102.95467293641	2206.74691110181\\
1803.31432580452	3215.65349486806\\
1767.42925685483	2646.78000160129\\
};
\addplot3[only marks,mark=asterisk,mark options={},mark size=5pt,color=green] plot table[row sep=crcr,]{%
2251.14726144461	3327.02225494686	300\\
};
\addplot[only marks,mark=diamond*,mark options={},mark size=3.6092pt,color=blue] plot table[row sep=crcr,]{%
959.194207575183	1251.171990357\\
2372.47106526323	1394.30747890552\\
263.011068279709	498.337998101655\\
624.861786845241	1736.9682990966\\
476.470571076389	827.845867787273\\
1924.37118987097	471.623403817135\\
2027.52366578952	789.177430179262\\
1528.73536653258	1492.82624142829\\
1464.32251151287	830.417524337401\\
1490.83469637876	1847.25739444395\\
937.079410959819	883.285140724237\\
410.617360335338	2008.13805127995\\
};
\addplot3[only marks,mark=asterisk,mark options={},mark size=5pt,color=blue] plot table[row sep=crcr,]{%
1385.26798535204	590.221976969278	300\\
};
\end{axis}
\end{tikzpicture}%

%% file: Fig2.tikz
\footnotesize
\begin{tikzpicture}
\begin{axis}[%
width=\figurewidth,
height=\figureheight,
scale only axis,
xmin=0,xmax=2,
xlabel={$R_r$},
xlabel near ticks,
ymin=0,ymax=80,
ylabel={Avg. \% of Unassociated SCs},
ylabel near ticks,
legend style={legend cell align=left,align=left,draw=white!15!black}
]
\addplot [color=black,solid,line width=1.5pt,mark size=4pt,mark=x,mark options={solid}]
  table[row sep=crcr]{%
0.1	77.0666666666667\\
0.2	62.8733333333333\\
0.3	51.9633333333333\\
0.4	42.6033333333333\\
0.5	34.2666666666667\\
0.6	26.9133333333333\\
0.7	20.0966666666667\\
0.8	13.59\\
0.9	7\\
1	0\\
1.2	0\\
1.4	0\\
1.6	0\\
1.8	0\\
2	0\\
};
\addlegendentry{B\&B};

\addplot [color=red,line width=1.5pt,mark size=3pt,only marks,mark=o,mark options={solid}]
  table[row sep=crcr]{%
0.1	77.4533333333333\\
0.2	63.9933333333333\\
0.3	53.3566666666667\\
0.4	44.1\\
0.5	36.02\\
0.6	28.77\\
0.7	21.74\\
0.8	14.44\\
0.9	7.52666666666667\\
1	0\\
1.2	0\\
1.4	0\\
1.6	0\\
1.8	0\\
2	0\\
};
\addlegendentry{\ref{Cent_Algo}};

\addplot [color=blue,line width=1.5pt,mark size=2.5pt,only marks,mark=asterisk,mark options={solid}]
  table[row sep=crcr]{%
0.1	77.4533333333333\\
0.2	63.9933333333333\\
0.3	53.3566666666667\\
0.4	44.1\\
0.5	36.02\\
0.6	28.77\\
0.7	21.74\\
0.8	14.44\\
0.9	7.52666666666667\\
1	0\\
1.2	0\\
1.4	0\\
1.6	0\\
1.8	0\\
2	0\\
};
\addlegendentry{\ref{Dist_Algo}};

\end{axis}
\end{tikzpicture}%

%% file: Fig3.tikz
\footnotesize
\begin{tikzpicture}
\begin{axis}[%
width=\figurewidth,
height=\figureheight,
scale only axis,
xmin=0,xmax=2,
xlabel={$R_b$},
xlabel near ticks,
ymin=0, ymax=60,
ylabel={Avg. \% of Unassociated SCs},
ylabel near ticks,
legend style={legend cell align=left,align=left,draw=white!15!black}
]
\addplot [color=black,solid,line width=1.5pt,mark size=4.0pt,mark=x,mark options={solid}]
  table[row sep=crcr]{%
0.1	50.8533333333333\\
0.2	31.34\\
0.3	18.9\\
0.4	10.3733333333333\\
0.5	4.97\\
0.6	2.04666666666667\\
0.7	0.753333333333333\\
0.8	0.293333333333333\\
0.9	0.113333333333333\\
1	0\\
1.2	0\\
1.4	0\\
1.6	0\\
1.8	0\\
2	0\\
};
\addlegendentry{B\&B};

\addplot [color=red,line width=1.5pt,mark size=3.0pt,only marks,mark=o,mark options={solid}]
  table[row sep=crcr]{%
0.1	56.34\\
0.2	35.45\\
0.3	21.9766666666667\\
0.4	13.1133333333333\\
0.5	7.24666666666667\\
0.6	3.70333333333333\\
0.7	1.77333333333333\\
0.8	0.88\\
0.9	0.38\\
1	0.17\\
1.2	0.0666666666666667\\
1.4	0.0133333333333333\\
1.6	0.00333333333333333\\
1.8	0\\
2	0\\
};
\addlegendentry{\ref{Cent_Algo}};

\addplot [color=blue,line width=1.5pt,mark size=2.5pt,only marks,mark=asterisk,mark options={solid}]
  table[row sep=crcr]{%
0.1	55.77\\
0.2	35.1433333333333\\
0.3	21.8666666666667\\
0.4	13.05\\
0.5	7.23333333333333\\
0.6	3.70666666666667\\
0.7	1.78\\
0.8	0.88\\
0.9	0.38\\
1	0.17\\
1.2	0.0666666666666667\\
1.4	0.0133333333333333\\
1.6	0.00333333333333333\\
1.8	0\\
2	0\\
};
\addlegendentry{\ref{Dist_Algo}};

\end{axis}
\end{tikzpicture}%

%% file: Fig4_new.tikz
\footnotesize
\begin{tikzpicture}
\begin{axis}[%
width=0.31748\figurewidth,
height=0.31748\figureheight,
at={(0.21\figurewidth,0.33\figureheight)},
scale only axis,
xmin=1, xmax=11,
xtick={1,3,5,7,9,11},
xticklabels={10,12,14,16,18,20},
ymin=0, ymax=3.5,
ytick={0,1,2,3},
yticklabels={0,1,2,3},
y filter/.code={\pgfmathparse{#1*1000}\pgfmathresult}
]
\addplot [color=red,line width=1.5pt,mark size=4pt,only marks,mark=o,mark options={solid},forget plot]
  table[row sep=crcr]{%
1	0.00318446888913064\\
2	0.00304612081440626\\
3	0.00327026932716446\\
4	0.00329407997503987\\
5	0.00328422729316039\\
6	0.0032296270144116\\
7	0.00323865863946779\\
8	0.00327232196922269\\
9	0.00337823829942711\\
10	0.00318528994595393\\
11	0.00325631136116853\\
};
\addplot [color=blue,line width=1.5pt,mark size=4pt,only marks,mark=x,mark options={solid},forget plot]
  table[row sep=crcr]{%
1	0.000390823047886087\\
2	0.000476623485919902\\
3	0.000412991582114919\\
4	0.00038055983759496\\
5	0.000512749986144666\\
6	0.000568581850128393\\
7	0.000349359678309937\\
8	0.000314054234908462\\
9	0.000442960156165008\\
10	0.000535329048785144\\
11	0.000397391502472407\\
};
\end{axis}

\begin{axis}[%
width=\figurewidth,
height=\figureheight,
scale only axis,
xmin=10, xmax=40,
xtick={10,15,20,25,30,35,40},
xticklabels={1,5,10,15,20,25,30},
xlabel near ticks,
xlabel={Scenarios},
ymin=-0.685945765091538, ymax=0029.2357336245348,
ylabel near ticks,
ylabel={Run Time (ms)},
legend style={at={(0.74,0.25)},anchor=south west,legend cell align=left,align=left,draw=white!15!black},
y filter/.code={\pgfmathparse{#1*1000}\pgfmathresult}
]
\addplot [color=black,line width=1.5pt,mark size=4pt,only marks,mark=square,mark options={solid}]
  table[row sep=crcr]{%
1	0.373256126623383\\
2	0.0477054540752129\\
3	0.024395650862007\\
4	0.0240298700472313\\
5	0.250551647553148\\
6	0.0249896854736574\\
7	0.0263961558119534\\
8	0.0231254759563773\\
9	0.0235253306293195\\
10	0.0277090256723942\\
11	0.0267602945130825\\
12	0.0230487071433996\\
13	0.0261937653050123\\
14	0.022940738171137\\
15	0.0230868862856826\\
16	0.0228545272046915\\
17	0.0227395792494309\\
18	0.0226262734078169\\
19	0.0228200428181133\\
20	0.0239756802968941\\
21	0.0253534136463749\\
22	0.0238812587622158\\
23	0.0227925374145331\\
24	0.0248193161828247\\
25	0.0243509032651377\\
26	0.0230105280011166\\
27	0.0240885756100965\\
28	0.023269160900453\\
29	0.0228972221595026\\
30	0.0229583908928377\\
31	0.0251136650539742\\
32	0.0248213688248829\\
33	0.0236378154141103\\
34	0.0241673970651324\\
35	0.0240586070360464\\
36	0.0225885047939455\\
37	0.022623810237347\\
38	0.0227757057496557\\
39	0.0282357336245348\\
40	0.0245578095846068\\
41	0.027333392175739\\
42	0.0242068077926503\\
43	0.0231476444906061\\
44	0.0229986226771789\\
45	0.0249732643371916\\
46	0.0235516044476648\\
47	0.0229538750803096\\
48	0.0226320208055799\\
49	0.0232864030937421\\
50	0.0232679293152181\\
51	0.0228594535456313\\
52	0.024422335208764\\
53	0.0243439242821398\\
54	0.0244001666745351\\
55	0.0227063264480877\\
56	0.0231677603827767\\
57	0.0227223370561418\\
58	0.0232137395648809\\
59	0.0240052383425326\\
60	0.0227941795281797\\
61	0.0231574971724856\\
62	0.0250290962011753\\
63	0.026099343770334\\
64	0.0254839616812781\\
65	0.0276269199900652\\
66	0.0227133054310856\\
67	0.0231460023769595\\
68	0.0226640420216882\\
69	0.0227334213232562\\
70	0.0230240754387009\\
71	0.0243238083899692\\
72	0.0251001176163899\\
73	0.139542712402269\\
74	0.024200239338064\\
75	0.0299529739704461\\
76	0.0259121428146238\\
77	0.023542162294197\\
78	0.0273789608294316\\
79	0.0241579549116645\\
80	0.0226903158400335\\
81	0.0249313904392038\\
82	0.0257779000240159\\
83	0.0241160810136768\\
84	0.0239165642056173\\
85	0.0229054327277355\\
86	0.0228504219205751\\
87	0.0242137867756483\\
88	0.0255915201251291\\
89	0.023540930708962\\
90	0.0233192453666737\\
91	0.0231698130248349\\
92	0.02402207000741\\
93	0.0226541893398087\\
94	0.023345519185019\\
95	0.0228122427782921\\
96	0.0257011312110383\\
97	0.0280682380325836\\
98	0.0225843995098291\\
99	0.0234699092937474\\
100	0.0227564109143084\\
};
\addlegendentry{B\&B};

\addplot [color=red,line width=1.5pt,mark size=4pt,only marks,mark=o,mark options={solid}]
  table[row sep=crcr]{%
1	0.0240368490302293\\
2	0.00474365579655854\\
3	0.00335073289584689\\
4	0.0034866178001014\\
5	0.0033252801343249\\
6	0.00321813221888554\\
7	0.00312494226944211\\
8	0.00332281696385503\\
9	0.00319473209942177\\
10	0.00323989022470273\\
11	0.00334703814014208\\
12	0.0031257633262654\\
13	0.00333513281620438\\
14	0.00321813221888554\\
15	0.00322018486094376\\
16	0.0033536065947284\\
17	0.00334826972537702\\
18	0.00321484799159238\\
19	0.00332979594685299\\
20	0.00318446888913064\\
21	0.00304612081440626\\
22	0.00327026932716446\\
23	0.00329407997503987\\
24	0.00328422729316039\\
25	0.0032296270144116\\
26	0.00323865863946779\\
27	0.00327232196922269\\
28	0.00337823829942711\\
29	0.00318528994595393\\
30	0.00325631136116853\\
31	0.00326493245781307\\
32	0.00323619546899792\\
33	0.00325466924752195\\
34	0.00326287981575485\\
35	0.00302518386541237\\
36	0.00303585760411514\\
37	0.00316640563901826\\
38	0.00358432356207292\\
39	0.0031758477924861\\
40	0.00327847989539736\\
41	0.00311139483185783\\
42	0.00334580655490715\\
43	0.00423706373658855\\
44	0.0033417012707907\\
45	0.00314916344572917\\
46	0.00297181517189851\\
47	0.00308963682604064\\
48	0.00324194286676095\\
49	0.00332651171955983\\
50	0.00325795347481511\\
51	0.00323947969629108\\
52	0.00323619546899792\\
53	0.00323167965646983\\
54	0.00344515443052525\\
55	0.00322346908823692\\
56	0.0034492597146417\\
57	0.0032460481508774\\
58	0.00334334338443728\\
59	0.00333390123096944\\
60	0.00322429014506021\\
61	0.00333964862873247\\
62	0.00326452192940143\\
63	0.00349277572627608\\
64	0.00328381676474875\\
65	0.00331091163991732\\
66	0.00318898470165874\\
67	0.00374566122784943\\
68	0.00325343766228701\\
69	0.00323824811105615\\
70	0.0031220685705606\\
71	0.00330229054327277\\
72	0.00320622689494783\\
73	0.00275792986943144\\
74	0.00323209018488147\\
75	0.00301984699606098\\
76	0.00316968986631142\\
77	0.00333841704349754\\
78	0.00288560420545305\\
79	0.0032542587191103\\
80	0.0033453960264955\\
81	0.00306746829181181\\
82	0.00302148910970756\\
83	0.00340040683365594\\
84	0.00328874310568849\\
85	0.00326903774192952\\
86	0.00333677492985096\\
87	0.00334457496967221\\
88	0.00326041664528498\\
89	0.00320294266765467\\
90	0.0032169006336506\\
91	0.00309743686586189\\
92	0.00331665903768035\\
93	0.00316886880948813\\
94	0.00324153233834931\\
95	0.00329038521933507\\
96	0.00323578494058628\\
97	0.00314957397414082\\
98	0.00319842685512658\\
99	0.00324071128152602\\
100	0.00328053253745559\\
};
\addlegendentry{CMCA};

\addplot [color=blue,line width=1.5pt,mark size=3.5pt,only marks,mark=x,mark options={solid}]
  table[row sep=crcr]{%
1	0.0147818965181033\\
2	0.00500228869589492\\
3	0.000443781212988298\\
4	0.000403549428647083\\
5	0.00302600492223566\\
6	0.000419149508289595\\
7	0.000404781013882018\\
8	0.000357570246542838\\
9	0.000376044025066865\\
10	0.000366601871599029\\
11	0.000460612877865745\\
12	0.000392054633121022\\
13	0.00156123954948612\\
14	0.0004602023494541\\
15	0.000430233775404011\\
16	0.000403549428647083\\
17	0.000338685939607165\\
18	0.000337864882783875\\
19	0.000423254792406045\\
20	0.000390823047886087\\
21	0.000476623485919902\\
22	0.000412991582114919\\
23	0.00038055983759496\\
24	0.000512749986144666\\
25	0.000568581850128393\\
26	0.000349359678309937\\
27	0.000314054234908462\\
28	0.000442960156165008\\
29	0.000535329048785144\\
30	0.000397391502472407\\
31	0.000403959957058728\\
32	0.000427360076522496\\
33	0.000331296428197554\\
34	0.000479907713213062\\
35	0.000406833655940243\\
36	0.000473749787038386\\
37	0.000534097463550209\\
38	0.000351412320368162\\
39	0.000368654513657254\\
40	0.000486065639387738\\
41	0.000537381690843369\\
42	0.000434339059520462\\
43	0.000414633695761499\\
44	0.000340738581665391\\
45	0.000396159917237472\\
46	0.000463076048335615\\
47	0.000450349667574619\\
48	0.000482370883682933\\
49	0.000339506996430455\\
50	0.000442960156165008\\
51	0.000454454951691069\\
52	0.000465539218805486\\
53	0.000528350065787178\\
54	0.00050659205996999\\
55	0.000347307036251711\\
56	0.000468823446098646\\
57	0.000370707155715479\\
58	0.000522602668024147\\
59	0.000354696547661322\\
60	0.00037481243983193\\
61	0.000349770206721582\\
62	0.00042038109352453\\
63	0.000387128292181281\\
64	0.000454865480102714\\
65	0.000537381690843369\\
66	0.000383023008064831\\
67	0.000375222968243575\\
68	0.000363728172717513\\
69	0.000323085859964654\\
70	0.000490581451915833\\
71	0.000578845060419519\\
72	0.000478676127978127\\
73	0.000502897304265185\\
74	0.000442960156165008\\
75	0.000444602269811588\\
76	0.000443370684576653\\
77	0.000373170326185349\\
78	0.000520139497554277\\
79	0.000432696945873882\\
80	0.000401907315000503\\
81	0.000466360275628776\\
82	0.000529171122610468\\
83	0.000471286616568516\\
84	0.000403959957058728\\
85	0.000319391104259848\\
86	0.000384254593299766\\
87	0.00037563349665522\\
88	0.000427360076522496\\
89	0.000449528610751329\\
90	0.000397391502472407\\
91	0.000497560434913799\\
92	0.000380970366006605\\
93	0.000496328849678864\\
94	0.000376044025066865\\
95	0.000416686337819725\\
96	0.000513981571379601\\
97	0.000454044423279424\\
98	0.000357159718131193\\
99	0.000429412718580721\\
100	0.000376044025066865\\
};
\addlegendentry{DMCA};

\end{axis}
\end{tikzpicture}%